\documentclass[twocolumn,preprintnumbers,amsmath,amssymb,aps]{revtex4}

\usepackage{graphicx}
\usepackage{amssymb}
\usepackage{epstopdf}
\usepackage{bm}

\begin{document}
    
\title{Testing Many-Worlds Quantum Theory By Measuring Pattern Convergence Rates}
    
\author{Frank  J. Tipler}
\affiliation{Department of Mathematics and Department of Physics, Tulane University, New Orleans, LA 70118}

\date{\today}

\begin{abstract}
The Born Interpretation of the wave function gives only the relative frequencies as the number of observations approaches infinity.  Using the Many-Worlds Interpretation of quantum mechanics, specifically the fact that there must exist other versions of ourselves in the multiverse, I show that the observed frequencies should approach the Born frequencies as $\sim 1/N$, where $N$ is the number of observations.  We can therefore test the central claim of the MWI by measuring the convergence rate to the final Born frequency.  Conversely, the MWI allows us to calculate this convergence rate.
\end{abstract}

\maketitle

A cat's eye has long been known to be sensitive to single photons \cite{Barlow1971}, and extensive recent work \cite{Field2005} indicates that this is true of all mammalian eyes, including human eyes.  In the past few years, enormous effort \cite{Holl2006} has been expended to develop CCDs  --- the central elements in ordinary digital cameras, and in all modern astronomical cameras ---  that are also sensitive to single photons.  But a central question has heretofore been ignored, namely, how rapidly do the laws of physics permit individual photons to add up and yield the final pattern?  This is a question which standard quantum mechanics cannot answer, but I shall show that Many-Worlds quantum theory can.  This greater scope of Many-Worlds theory allows a direct test of Many-Worlds quantum mechanics, and shows not only that the Many-Worlds idea is testable --- an issue debated in an issue of {\it Nature} last summer \cite{Buchanan2007}, \cite{Tegmark2007} on the idea --- but also that it is an essential idea for use in many areas of biology, physics, and engineering.

Non-Many-Worlds quantum mechanics, based on the Born Interpretation of the wave function, gives only relative frequencies asymptotically as the number of observations goes to infinity.  In actual measurements, the Born frequencies are seen to gradually build up as the number of measurements increases, but standard theory gives no way to compute the rate of convergence.  The Many-Worlds Interpretation \cite{Everett1957} allows such a computation: the MWI says the absolute difference between the observed distribution and the Born frequencies decreases inversely as the number $N$ of measurements.

Such a rate of convergence has long been suspected \cite{DeWittGraham1973}, but earlier proofs are known to be defective, and have not been stated in a way that allows an experimental test.  I shall state the simple, {\bf EASILY TESTABLE FORMULA} here, and publish the proof elsewhere.  The method I use to obtain this formula can with equal ease be applied to other cases, like the human eye, and the CCD camera.

An outline of the proof is as follows.  Many-Worlds quantum mechanics asserts that before measurements, identical copies of the observer exist in parallel universes.  Bayesian probability theory \cite{Jaynes03}, applied to identical observers coupled to the wave function of the quantum system  being observed, yields a Bayesian probability density (which in Bayesian theory is NOT a relative frequency density) equal to $|\psi|^2$.  Standard Bayesian analysis then yields the rate of convergence of the observations to $|\psi|^2$. 

The theorem stated here is analogous to the Berry-Esseen Theorem \cite{Petrov} which is an unpacking of the Central Limit Theorem.  The Berry-Esseen Theorem states how fast any probability distribution with finite second moment must approach a Gaussian distribution as $N \rightarrow +\infty$.  

To state the quantum mechanical version of the  Berry-Esseen Theorem, note that for a one dimensional system, the Born distribution defines a {\it cumulative distribution function,} (cdf), namely $F_B(x) = \int_{-\infty}^x \psi^*(t)\psi(t)\,dt$.  Consider the measurement of the frequency distribution of photons or electrons incident on a screen after passing through a single (or double) slit.  The distribution will depend on one variable: the distance along the screen, call it $x$, with $x=0$ the location of the central peak.  The variable $x \in (-\infty, +\infty)$, so divide up this region into $M+ 2$ bins; one of size $(-\infty, m_-)$, one of size $(m_+, +\infty)$, and $M$ bins of equal size $\Delta\ell$.  The numbers $m_-$ and $m_+$ are determined by the condition that there are no observed particles in either region $(-\infty, m_-)$ or in region $(m_+, +\infty)$.  

Let $N$ be the total number of particles observed, and let $N_i$ be the number of particles observed to be in the $ith$ bin.  We observe a pattern on the screen after $N$ particle observations.  Let the Born cdf be normalized as usual so that if $x= +\infty$, the integral equals one.   The normalized measured cdf is the ratio $(\sum_{i=1}^{j(x)} N_i\Delta\ell)/(\sum_{i=1}^M N_i\Delta\ell)$, where $j(x)$ is the $jth$ bin, chosen so that the upper end of the $jth$ bin is in position $x$ on the real line.  The factor $\Delta\ell$ cancels out, so we can state the formula for the Many-Worlds prediction for how rapidly the observed frequencies will approach the Born relative frequencies as

\begin{equation}
\left|\frac{\sum_{i=1}^{j(x)} N_i}{N} - \int_{-\infty}^x \psi^*(t)\psi(t)\,dt \right| \leq \frac{C(M,e)}{N}
\label{eq:formula}
\end{equation}	

\noindent
where $C(M.e)$ is a constant that will depend on the number of bins $M$, and on the detector efficiency $e$.  As in the Berry-Esseen Theorem, the important fact is that the LHS of (\ref{eq:formula}) will be independent of $x$, and the approach of the two terms in (\ref{eq:formula}) to each other will be $\sim 1/N$.  The Berry-Esseen rate is $\sim 1/\sqrt N$, slower than (\ref{eq:formula}).  The experimental strategy will be to record the locations of the particles as they are detected one by one, and after all the data are taken, choose the bin size, the number of bins, and the numbers $m_-$ and $m_+$.  Then see if there is a constant such that (\ref{eq:formula}) holds as $N$ is increased.  It is obvious that (\ref{eq:formula}) holds in two extreme cases, namely  (i) $N=1$, and (ii) $x = +\infty$.

The important point is that by testing (\ref{eq:formula}), the rate of convergence to the Born distribution can be experimentally investigated.  A further point is that (\ref{eq:formula}) is derived assuming the actual existence of many versions of human observers out in the Many-Worlds.  Thus testing the formula tests for the presence of our analogues in the Many-Worlds.  As one watches the distribution (\ref{eq:formula}) build up, one is really watching the activity of other versions of oneself in the Many-Worlds, just as seeing the Sun set is really seeing the Earth rotate.

\end{document}